# Improvement of the electrocaloric effect and energy storage performances in Pb-free ferroelectric $Ba_{0.9}Sr_{0.1}Ti_{0.9}Sn_{0.1}O_3$ ceramic near room temperature.

S. Khardazi[1], H. Zaitouni[1], S. Belkhadir[1], D. Mezzane[1,2], M. Amjoud[1], Y. Gagou[2], B. Asbani[2], I. Lukyanchuk[2,3], S. Terenchuk[3]

[1] IMED-Lab, Cadi-Ayyad University, Faculty of Sciences and Technology, Department of Applied Physics, Marrakech, Morocco

[2] Laboratory of Physics of Condensed Matter (LPMC), University of Picardie Jules Verne, Scientific Pole, 33 Rue Saint-Leu, Amiens Cedex 1 80039, France

[3] Kiev National University of Construction and Architecture, Faculty of Automation and Information Technologies, Department of Information Design Technologies and Applied Mathematics, Vozdukhoflotsky Avenue, 31, 03037 Kiev, Ukraine

## Abstract

Perovskite-type $Ba_{0.9}Sr_{0.1}Ti_{0.9}Sn_{0.1}O_3$(BSTSn) ceramic was synthesized by the sol-gel method. The P-E hysteresis loops were recorded at different temperatures to investigate the ferroelectric and energy storage properties of BSTSn ceramic. Enhanced recoverable energy density and high energy storage efficiency were found to be 58.08 mJ/cm$^3$ and 84.36 %, respectively at room temperature (RT) under a moderate applied electric field of 20 kV/cm. The electrocaloric effect (ECE) in the BSTSn ceramic was explored using two different approaches based on P – E hysteresis loops, and pyroelectric current measurements. The largest electrocaloric (EC) temperature change, $\Delta T_{max} \approx 0.63$ K was determined using the Maxwell relationship obtained near RT under 20 kV/cm. The corresponding EC responsivity ($\xi_{max}$) value of 0.31 K·mm/kV is one of the highest reported values in lead-free ferroelectrics near RT. This study demonstrates that the BSTSn ceramic is a promising candidate for solid-state cooling and energy storage applications.

**Keywords**: Lead-free ferroelectric, sol-gel, electrocaloric effect, energy storage.

## 1. Introduction

In the last few decades, energy storage has become one of the vital topics in the scientific research, because it consists of preserving an energy generated for later use [1]. With the excessive consumption of traditional resources like fossil fuels, the acceleration of environmental pollution, and the extensive demand of the electronics industry, the development of new dielectric materials for ecological energy storage has become a mandatory challenge [2]–[5]. Recently, dielectric capacitors have attracted a great attention due to their highest power density compared to other energy storage devices like fuel cells, batteries, and super-capacitors because of the fast charge-discharge rate [6]–[9]. Therefore, they are appealing candidates for advanced pulsed power systems such as high-power microwaves, electromagnetic devices, and hybrid electric vehicles (HEVs) [10]. Extensive research has revealed that dielectric materials with moderate dielectric breakdown strength (BDS), high maximum polarization ($P_{max}$), and low remnant polarization ($P_r$) are required for improving of the capacitor efficiency [11], [12]. Among these materials, antiferroelectric (AFEs) ceramics are thought to be particularly promising materials for energy storage applications. Unfortunately, however that the most studied anti-ferroelectrics are lead-based materials like $PbZrO_3$ (PZ) [13], $PbZr_{1-x}Ti_xO_3$ (PZT) [14] or, $Pb_{1-x}La_x(Zr_{1-y}Ti_y)_{1-x/4}O_3$ (PLZT) [15], that are harmful to the environment, limiting their applications [16], [17].

On the other hand, electrocaloric refrigeration has been regarded as a promising future candidate for vapor-cycle refrigeration [18]–[20]. The electrocaloric effect (ECE), which is the inverse process of the pyroelectric effect, is defined as the temperature change and isothermal entropy change of a dielectric material caused by the application or removal of an external electric field [21], [22].

The lead-free $BaTiO_3$ (BT) ceramic with relatively high dielectric permittivity and low loss features is a promising eco-friendly material for energy storage applications [4], [23]. Recently, BT has also been recognized as a potentially promising lead-free ceramic for future solid-state electrocaloric cooling applications [24]–[26]. However, the low breakdown field strength and high hysteresis loss of these ceramics are significant drawbacks that limit their energy storage applications [27], [28]. Hence, doping with other metal oxides or generating solid solutions with other compounds can be used to enhance the energy storage properties of BT as well as other properties at the near room temperatures (RT) [29]. The study of energy storage and ECE in lead-free $BaTiO_3$ derivative system, such as $(Ba_{1-x}Sr_x)TiO_3$ solid solutions has revealed intriguing properties [30], [31]. Peng-Zu *et al*. [31] obtained an energy storage density of 0.2812 J/cm$^3$and an adiabatic temperature change $\Delta T_{max}$ = 0.49 K under an electric field of 75 kV/cm in $Ba_{0.65}Sr_{0.35}TiO_3$ ceramics. Qian *et al*. [32] also reported a high energy storage density of 1.081 J/cm$^3$ in $Ba_{0.4}Sr_{0.6}TiO_3$ ceramic under a high electric field of 167kV/cm. In another lead-free BT-based material with enhanced EC effect properties [33] Merselmiz *et al*. [34] have found a value of $\Delta T_{max}$ = 0.71K at 49°C under an electric field of 25kV/cm. Sanlialp *et al* [35] reported a value of $\Delta T_{max}$ = 0.63 K under an electric field of 20kV/cm. It is worthing to note that many researchers focus mostly on the improvement of electrocaloric temperature and/or energy storage density for bulk materials, neglecting the practical applications in room temperature. To the best of our knowledge, only a few studies have reported on the energy storage and electrocaloric properties of $Ba_{1-x}Sr_xTi_{1-y}Sn_yO_3$ solid solutions, which are obtained by replacing Ba ions with Sr ions and Ti ions with Sn ions [36]–[38]. Indeed, our previous research [39] highlighted that the partial replacement of $Sr^{2+}$and $Sn^{4+}$ in BT matrix (with nominal composition $Ba_{0.9}Sr_{0.1}Ti_{0.9}Sn_{0.1}O_3$) has shifted the ferroelectric paraelectric (FE – PE) phase transition to RT and enhanced its dielectric properties. Therefore, the current paper focuses on the Pb-free $Ba_{0.9}Sr_{0.1}Ti_{0.9}Sn_{0.1}O_3$ (named as BSTSn) ceramic elaborated by sol-gel prossess as possible candidate for energy storage and solid-state refrigeration applications near RT.

## 2. Experimental details

The polycrystalline ceramic $Ba_{0.9}Sr_{0.1}Ti_{0.9}Sn_{0.1}O_3$ (BSTSn) with a thickness of 0.69 mm was prepared via sol-gel method. A description of synthesis method has been reported elsewhere [39]. The obtained powder was calcined at 1000°C for 5 h. Then, the pellet was then sintered at 1420 °C for 5 hours after being pressed at 2.5 ton/cm$^2$. The X-ray diffraction (XRD) pattern was recorded at room temperature using the Panalytical X-Pert Pro under Cu-Kα radiation with λ ~ 1.540598 Å. The grain morphology of the sintered ceramic was observed by using the TESCAN VEGA3 Scanning Electron Microscope (SEM). Raman spectra was recorded using a micro-Raman RENISHAW with a CCD detector and green laser excitation of 532 nm. Polarization versus electric field (P – E) hysteresis loops were measured at 200 Hz using a ferroelectric test system (PolyK Technologies State College, PA, USA) for different temperatures. Pyroelectric current was determined using a Keithley 2635B Source Meter.

## 3. Results and discussion

3.1. X-ray diffraction and microstructural analysis

Room temperature X-ray diffraction (XRD) patterns of the BSTSn powder are shown in Fig. 1. It can be seen that the sample shows a pure perovskite structure without any secondary phase, indicating that Sr and Sn have successfully diffused into BT lattice and formed a solid solution. The Rietveld refinement of the XRD data was carried out using Fullprof Suite software, revealing a tetragonal phase with P*4mm* space group. The refined parameters have been reported in our earlier work [39].

The SEM micrograph of BSTSn ceramic sintered at 1420°C/ 5h is displayed in Fig. 2a. The sample shows a dense microstructure with an inhomogeneous grain size distribution. The obtained grain size of 6.80 ± 0.17 μm was determined using the Gaussian distribution as shown in Fig. 2b. The smaller resulting grains size indicates the good quality of the obtained ceramic by sol-gel method, which is very suitable for electrical and energy storage applications [40], [41].

**(Insert Figure 1, here)**

**(Insert Figure 2, here)**

3.2. Raman spectroscopy study

Due to its high sensitivity to crystal lattice vibration and phase transition, Raman spectroscopy is a powerful analytical tool for investigating the ferroelectric phase transition and local structure. Fig. 3a displays the thermal evolution of the Raman spectra of BSTSn ceramic. All of these Raman active modes are common in the tetragonal BT Raman spectra, as reported elsewhere [42], [43]. In the Raman spectrum recorded at -20 °C, the band situated at 191 $cm^{-1}$ is assigned to $A_1$(TO), the sharp peak at 303$cm^{-1}$ is ascribed to the combined mode $B_1$, E(TO, LO), and the band at 514 $cm^{-1}$ is attributed to the $A_1$(TO) mode, while the band at 723 $cm^{-1}$ is assigned to the phonon mode of low-intensity $A_1$/E(LO). In addition, the presence of an interference effect around 170 $cm^{-1}$, which occurs only in the tetragonal phase [37], [38] confirms the tetragonal symmetry of BSTSn bulk ceramic at -20 °C. The strong peak near 191 $cm^{-1}$ is a feature of the Rhombohedral phase [37], [44]. By increasing temperature from -20 to 25 °C, the sharpness of the peak near 303 $cm^{-1}$and the dip at 170 $cm^{-1}$ vanishes and other peaks broaden, revealing a decrease of the tetragonality in BSTSn ceramic. On further increasing the temperature above 30°C, the bands at 303 $cm^{-1}$ and at 723 $cm^{-1}$ disappear, while those located at 514 $cm^{-1}$and ~191 – 232 $cm^{-1}$ become flat and wide, indicating that the phase structure has changed to the paraelectric cubic phase.

Fig. 3b depicts the variation of the full width at half maxima (FWHM) of the band centered at 514 $cm^{-1}$ as a function of the temperature. One can see that a shift in the slope occurs around 25 °C that is associated with the FE – PE phase transition, which is well in accordance with the phase transition in our BSTSn at 25°C determined by the dielectric measurements [39].

**(Insert Figure 3, here)**

3.3. Ferroelectric and energy storage properties

Fig. 4a shows temperature-dependent P–E hysteresis loops of BSTSn ceramic recorded at 200 Hz and 20 kV/cm. The BSTSn ceramic displays typical and slimmer ferroelectric hysteresis curves that are changed to linear loops with rising temperature, confirming the (FE – PE) phase transition. The maximal polarization ($P_{max}$), remnant polarization ($P_r$) and coercive field ($E_c$) are found to be 11.22 μC/$cm^2$, 4.48 μC/$cm^2$ and 0.73 kV/cm, respectively at RT. Furthermore, as the temperature increases, $P_{max}$, $P_r$ and $E_c$ decrease gradually (Fig. 4b), and the P–E hysteresis loops become slimmer, indicating the appearance of the paraelectric phase.

**(Insert Figure 4, here)**

To investigate the energy storage properties, the P–E hysteresis loops of the BSTSn sample are recorded in the temperature range [25 – 105 °C] under an electrical field of 20kV/cm, as shown in Fig. 4a. In general, the electrical energy storage capacities of dielectric materials can be evaluated using their P–E hysteresis loops. The energy storage density can be easily obtained by integrating the area between the polarization axis and P–E curves discharge as depicted in Fig. 5. Thus, the recoverable energy density ($W_{rec}$), the total energy density ($W_{tot}$) and the energy storage efficiency ($\eta$) could be given as follow:

$$W_{tot}=\int_0^{Pmax} E\ dP \tag{1}$$

(2)

$$W_{rec} = \int_{Pr}^{Pmax} EdP$$

$$\eta = \frac{W_{rec}}{W_{tot}} * 100 = \frac{W_{rec}}{W_{rec}+W_{loss}} * 100 \qquad (3)$$

**(Insert Figure 5, here)**

Fig. 6 depicts the thermal evolution of $W_{rec}$, $W_{tot}$ and $\eta$ of BSTSn ceramic. It can be seen that the BSTSn sample has a maximum value of $W_{rec}$ near the Curie temperature (Tc), in accordance with dielectric measurements shown in our previous work [39]. It is interesting to note that, the highest values of $W_{rec}$ and $W_{tot}$ are found near room temperature as 58.08mJ/cm$^3$ and 68.84mJ/cm$^3$ respectively, with an energy efficiency of 84.36% at a relatively low applied electric field of 20 kV/cm. Moreover, the energy efficiency increases abruptly with the increase of the temperature from RT to 45 °C and then increases slowly to attain a maximum value of 92.87 % at 105 °C. Undoubtedly, a low coercive field and small remnant polarization result in a slimmer hysteresis loop, which also leads to a lower energy loss density and therefore a high storage efficiency. As a result, $\eta$ rises to a maximum at 105 °C, where the P–E hysteresis curves are linear (Fig. 4a). These improved results could be attributed to the small grain size and dense microstructure [41]. To situate our finding to literature, Table 1 displays a comparison of our data with other lead-free ceramics. Near RT, our BSTSn ceramic exhibits a high energy storage efficiency and moderate values of $W_{rec}$ and $W_{tot}$ compared to other ceramics due to the low applied electric field. The highest energy storage efficiency near RT could be attributed to the high polarization difference ($\Delta P = P_{max} - P_r$) and the low value of Ec. Recently, Merselmiz *et al.* [34] recorded an energy density of 85.1 mJ/cm$^3$ at 30 °C in BTSn ceramic synthesized by the solid-state method under a moderate electric field of 25 kV/cm. Furthermore, under a high electric field of 300 kV/cm, Huang *et al.* [45] have found at RT a high energy storage performances ($W_{rec}$= 1500 mJ/cm$^3$ and $\eta$ = 88.5 %) in 95wt%$Ba_{0.4}Sr_{0.6}TiO_3$–5wt%MgO composite prepared by the solid-state reaction. At high temperature (120 °C), Hanani *et al.* [46] have found a $W_{tot}$ and $W_{rec}$ of 16.5 and 14 mJ/cm$^3$ respectively, with an efficiency of 80% in BCZT ceramic. Later, Belkhadir *et al.*[47] reported an efficiency of 81.65% associated with $W_{tot}$ and $W_{rec}$ of 23.2 and 19 mJ/cm$^3$ at 140 °C under a low electric field of 12 kV/cm, in BCZST ceramics prepared by sol-gel method.

In practical applications, in addition to high energy density, excellent temperature stability is also required, which can be determined using the following equation [48]:

$$\frac{\Delta W_{rec,T}}{W_{rec,300K}} = \left|\frac{W_{rec,T} - W_{rec,300K}}{W_{rec,300K}}\right| \quad (4)$$

here, $W_{rec,T}$ is the $W_{rec}$ value at a given temperature, and $\Delta W_{rec,T}$ is the difference of $W_{rec,T}$ and $W_{rec,300\ K}$. The obtained value of $\Delta W_{rec,T}/W_{rec,300K}$ for BSTSn ceramic is less than 10% ($W_{rec}$ ~ 38.34 – 41.37 mJ/cm$^3$), indicating good thermal stability in the temperature range of 27 – 77 °C. The combination of these results may provide a promising way to develop efficient energy storage materials.

**(Insert Figure 6, here)**

**(Insert Table 1, here)**

3.4. Electrocaloric properties

To compute the ECE in eco-friendly BSTSn ceramic, the indirect Maxwell approach is employed. In this method, P–E hysteresis curves were recorded at 200 Hz at different temperatures (Fig. 4a). The ferroelectric order parameter P(E, T) is extracted from the upper branches of the corresponding P–E hysteresis loops. Assuming Maxwell relation $\left(\frac{\partial P}{\partial T}\right)_E = \left(\frac{\partial S}{\partial E}\right)_T$, the EC temperature change and the isothermal entropy change (ΔS) of an EC material can be calculated using the following equations [50]:

$$\Delta T = -\frac{1}{\rho}\int_{E_1}^{E_2} \frac{T}{C_p}\left(\frac{\partial P}{\partial T}\right)_E dE, \quad (5)$$

$$\Delta S = -\frac{1}{\rho}\int_{E_1}^{E_2} \left(\frac{\partial P}{\partial T}\right)_E dE, \quad (6)$$

here, ρ denotes the density of the BSTSn ceramic, which was found to be 5.47 g/cm$^3$, as reported in our previous work [39]. $C_p$ stands for the specific heat capacity, which is assumed to be 400 Jkg$^{-1}$K$^{-1}$ for BSTSn ceramic over the studied range temperature, as is frequently used for Sn doped BT in the literature [51]. $E_1$ and $E_2$ are the starting and final electric field limits, respectively. In this work, values of the critical factor $\left(\frac{\partial P}{\partial T}\right)_E$ were determined from seventh-order polynomial fits of P(T) data.

Fig. 7a depicts the thermal evolution of the adiabatic temperature change (ΔT) of BSTSn ceramic under different electric fields. It is worth noting that BSTSn ceramic exhibits a maximum value of ΔT around RT, which corresponds to the FE – PE phase transition at Curie temperature, as earlier discussed in our previous work [39]. In addition, ΔT increases with increasing the electric field since ΔT is directly proportional to the change in the electric field (dE) as shown in the Maxwell's equation (Eq. 5). For practical applications, the electrocaloric material should exhibit a high EC response and a large temperature change in response to a weak electric field. In this work, the BSTSn ceramic exhibits a high ECE temperature change of 0.63K under relatively low applied electric field of 20 kV/cm. The electrocaloric strength or the responsivity, defined as $\xi_{max} = (\Delta T_{max}/\ \Delta E_{max})$, is taken as a critical parameter to compare the ECE in different materials [52]. The corresponding EC responsivity in our BSTSn is found to be about 0.31 K·mm/kV near RT, which is one of the highest EC response values reported for BT-lead-free based materials near RT. Table 2 lists the BSTSn ceramic's EC response in comparison to previously published data for lead-free ferroelectrics materials. For instance, Zaitouni *et al.* [53] reported a value of 0.27 K·mm/kV at 340 K in $Ba_{0.9}Sr_{0.1}Ti_{0.95}Sn_{0.05}O_3$ Pb-free ceramic synthesized by a semi-wet method and Qi *et al*. [54] found a value of 0.275 K·mm/kV at 325 Kin $Ba_{0.94}Sr_{0.06}Ti_{0.9}Sn_{0.1}O_3$ ceramic prepared by solid-state method. Fig. 7b depicts the isothermal entropy change of BSTSn ceramic as a function of temperature under different electric fields. One can note that a similar pattern to ΔT is observed and the highest entropy change is found to be 0.7 J/Kg K near RT under an applied electric field of 20kV/cm. An important criteria to estimate the refrigeration cycle performance and to evaluate the material's efficiency is the coefficient of performance(COP), given by the following equation [55]:

$$COP = \frac{|Q|}{|W_{rec}|} = \frac{|\Delta S \times T|}{|W_{rec}|} \tag{7}$$

where Q and $W_{rec}$ are the isothermal heat and recovered density, respectively. Our results reveal a maximum COP value of 25 near the FE – PE phase transition (~ RT), which is much higher than several previous works of lead-free and lead-based materials (Table 2). For example, Hanani *et al.* [56] reported a value of 6.29 at 365 K in lead-free BCZT ceramics and Zhao *et al.* [57] obtained a COP value of 18 in $Pb_{0.9}7La_{0.02}(Zr_{0.75}Sn_{0.18}Ti_{0.07})O_3$ thick film. Large COP values indicate high cooling performance, implying that BSTSn ceramic with sizeable COP,

ΔT and ΔS values, have potential applications in future eco-friendly electrocaloric cooling technologies.

**(Insert Figure 7, here)**

**(Insert Table 2, here)**

In the second approach, instead of being computed from P(E, T), the derivative $\left(\frac{\partial P}{\partial T}\right)_E$(named pyroelectric coefficient), corresponding to the polarization shift caused by a temperature change, might be directly determined using pyroelectric current measurements.

The following equation (Eq. 8) presents the relationship between the pyroelectric current *i* and the pyroelectric coefficient $\left(\frac{\partial P}{\partial T}\right)$ [63]:

$$i = A\frac{\partial P}{\partial t} = A\frac{\partial P}{\partial T}\frac{\partial T}{\partial t} = A\frac{\partial P}{\partial T}\,r \quad (8)$$

Where *A* represents the electrode area, $\left(\frac{\partial P}{\partial T}\right)$ is the pyroelectric coefficient measured after field cooling at zero applied electric field and $r = \frac{\partial T}{\partial t}$ is the heating rate.

Fig. 8 displays the evolution of the pyroelectric coefficient as a function of temperature for BSTSn ceramic under a heating rate of 30°C/min. The highest pyroelectric coefficient value is found to be 150.31μC/m$^2$ K near RT.

In order to compute the adiabatic temperature change ΔT from experimental pyroelectric current measurements, the following equation is used [58]:

$$\Delta T = -\frac{T}{r\,\rho\,A\,Cp}\,i(t)\Delta E \quad (9)$$

Here, $E_1 = 0$ and $E_2$ represent the field applied to polarize the sample.

The EC responsivity of BSTSn ceramic can be determined using the following equation:

$$\frac{\Delta T}{\Delta E} = -\frac{Ti(t)}{r\rho ACp} \quad (10)$$

**(Insert Figure 8, here)**

Fig. 9 depicts a comparison of curves for the thermal evolution of electrocaloric temperature change ΔT obtained from P–E hysteresis loop analysis and pyroelectric measurements for BSTSn ceramic at 4 kV/cm. It should be noted that the selected applied electric field value was chosen because of the limited applied voltage of the experimental device (4 kV/cm, taking account the thickness of the sample). Both approaches follow the same pattern, with maximum EC responses values near RT, corresponding to the FE – PE phase transition. Moreover, comparable EC response (ΔT) is obtained with a slightly high value in the case of Maxwell's approach (0.12 vs 0.10 K with an error of 16%). This slight shift in the ΔT profile is usually attributed to the experimental error and numerical anomalies caused by the calculation of derivatives [58], [64].

**(Insert Figure 9, here)**

## Conclusion

In this work, the temperature dependence of ferroelectric properties of BSTSn ceramic synthesized by the sol-gel method are studied. The investigation of energy storage performances of BSTSn ceramic demonstrated the improved values of $W_{tot}$ = 68.84 mJ/ $cm^3$, and $W_{rec}$ = 58.08 mJ/ $cm^3$. The high $\eta$ = 84.36% are obtained near the RT under a low electric field of 20kV/cm. Using the indirect Maxwell's approach, the high EC response, demonstrating the ΔT = 0.63K and $\xi_{max}$ = 0.31 K·mm/kV with a high coefficient of performance exceeding 25 near RT is achieved at relatively small electric field of 20 kV/cm. These results are consistent with those performed by using pyroelectric measurements. We conclude that the BSTSn ceramic could be a promising material for environmentally friendly refrigeration applications as well as for high-efficiency energy storage applications near the RT.

## Acknowledgments

The authors gratefully acknowledge the financial support of CNRST Priority Program PPR15/2015 and the European H2020-MSCA-RISE-2017-ENGIMA action.

**Table captions**

Table 1: Comparison of the energy storage properties of BSTSn with other lead-free ceramics.

Table 2: Comparison of ECE values of BSTSn with other lead-free ferroelectric ceramics.

**Figure captions**

Figure 1: XRD patterns of $Ba_{0.9}Sr_{0.1}Ti_{0.9}Sn_{0.1}O_3$ powder calcined at 1000°C/5h.

Figure 2: (a) SEM micrograph and (b) grain size distribution of BSTSn ceramic sintered at 1420°C/5h.

Figure 3: Temperature dependence of (a) Raman spectra and (b) full width at half maximum (FWHM) for $A_1$(TO) band centered at 514 $cm^{-1}$ for BSTSn ceramic.

Figure 4: Thermal variation of (a)P – E hysteresis loops and (b) $P_{max}$, $P_r$ for BSTSn ceramic.

Figure 5: Schematic presentation of energy storage determined via P-E loops for BSTSn ceramic at 50 °C.

Figure 6: The evolution of energy storage parameters of BSTSn ceramic as a function of temperature.

Figure 7: Temperature-dependence of (a) EC temperature change $\Delta T$ and (b) isothermal entropy change $\Delta S$ under different applied electric fields in BSTSn ceramic.

Figure 8: Temperature- dependence of pyroelectric coefficient in BSTSn ceramic.

Figure 9: The electrocaloric temperature change $\Delta T$ versus temperature calculated from Maxwell relation and pyroelectric current measurements under 4kV/cm for BSTSn ceramic.

**Table 1**

| Samples | $W_{tot}$(mJ/cm$^3$) | $W_{rec}$(mJ/cm$^3$) | E(kV/cm) | $\eta$(%) | T(°C) | Refs. |
|---|---|---|---|---|---|---|
| **$Ba_{0.9}Sr_{0.1}Ti_{0.9}Sn_{0.1}O_3$** | **68.84** | **58.08** | **20** | **84.36** | **45** | **This work** |
| **$Ba_{0.9}Sr_{0.1}Ti_{0.9}Sn_{0.1}O_3$** | **29.74** | **27.62** | **20** | **92.87** | **105** | **This work** |
| $Ba_{0.85}Ca_{0.15}Zr_{0.08}Sn_{0.02}Ti_{0.9}O_3$ | 23.2 | 19 | 12 | *81.65* | 140 | [47] |
| $Ba_{0.85}Ca_{0.15}Zr_{0.10}Ti_{0.90}O_3$ | 16.5 | 14 | 6.5 | *80* | 129 | [46] |
| $BaZr_{0.05}Ti_{0.95}O_3$ | 302 | 218 | 50 | 72 | RT | [49] |
| $Ba_{0.65}Sr_{0.35}TiO_3$ | - | 281.2 | 75 | 75 | RT | [31] |
| $BaTi_{0.89}Sn_{0.11}O_3$ | 85.1 | 72.4 | 25 | 85.07 | 30 | [34] |
| 95 wt%$Ba_{0.4}Sr_{0.6}TiO_3$–5 wt%MgO | - | 1500 | 300 | 88.5 | RT | [45] |

**Table 2**

| Samples | $T_{EC}$(K) | ΔT(K) | $\xi_{max}$(K·mm/kV) | E(kV/cm) | Max COP | Refs. |
|---|---|---|---|---|---|---|
| **$Ba_{0.9}Sr_{0.1}Ti_{0.9}Sn_{0.1}O_3$** | **311** | **0.63** | **0.31** | **20** | **25** | **This work** |
| $Ba_{0.9}Sr_{0.1}Ti_{0.95}Sn_{0.05}O_3$ | 340 | 0.1886 | 0.27 | 7 | - | [53] |
| $Ba_{0.94}Sr_{0.06}Ti_{0.9}Sn_{0.1}O_3$ | 325 | 0.55 | 0.275 | 20 | - | [54] |
| $Ba_{0.95}Ca_{0.05}Zr_{0.1}Ti_{0.9}O_3$ | 374 | 0.20 | 0.25 | 8 | - | [58] |
| $Ba_{0.95}Ca_{0.05}Zr_{0.1}Ti_{0.9}O_3$ | 363 | 0.986 | 0.246 | 40 | 6.29 | [56] |
| $Ba_{0.85}Ca_{0.15}Ti_{0.9}Zr_{0.06}Sn_{0.04}O_3$ | 342 | 0.27 | 0.225 | 12 | - | [59] |
| $Ba_{0.65}Sr_{0.35}TiO_3$ | 296 | 0.42 | 0.21 | 20 | - | [60] |
| $BaSn_{0.1}Ti_{0.9}O_3$ | 343 | 0.4 | 0.20 | 20 | - | [33] |
| $Ba_{0.98}Ca_{0.02}Ti_{0.98}Li_{0.0067}Nb_{0.0133}O_{2.9967}$ | 374 | 0.11 | 0.14 | 8 | - | [61] |
| $0.94(Na_{0.5}Bi_{0.5})TiO_3$-$0.06BaTiO_3$ | 363 | 1.5 | 0.30 | 50 | - | [62] |

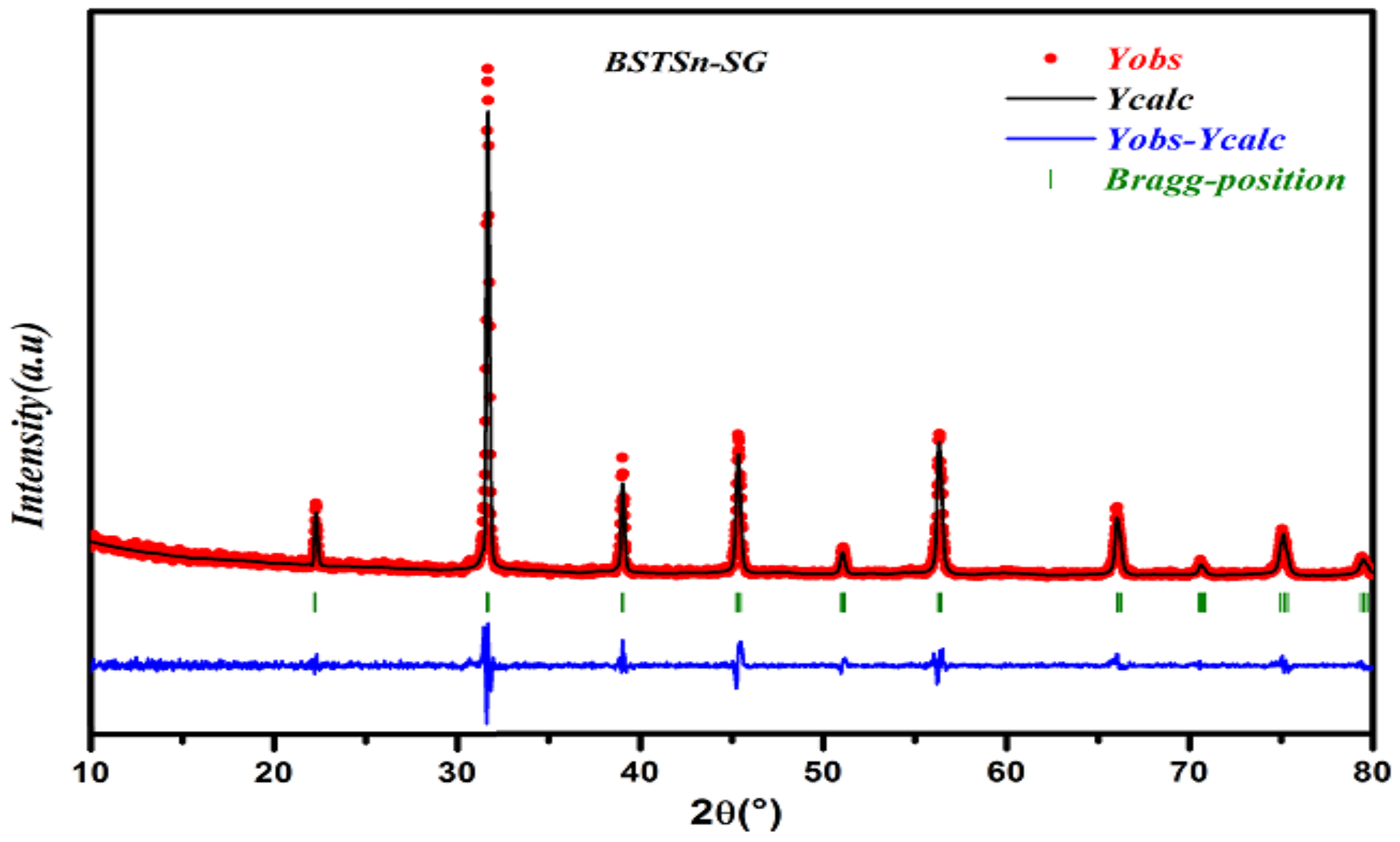

Figure 1

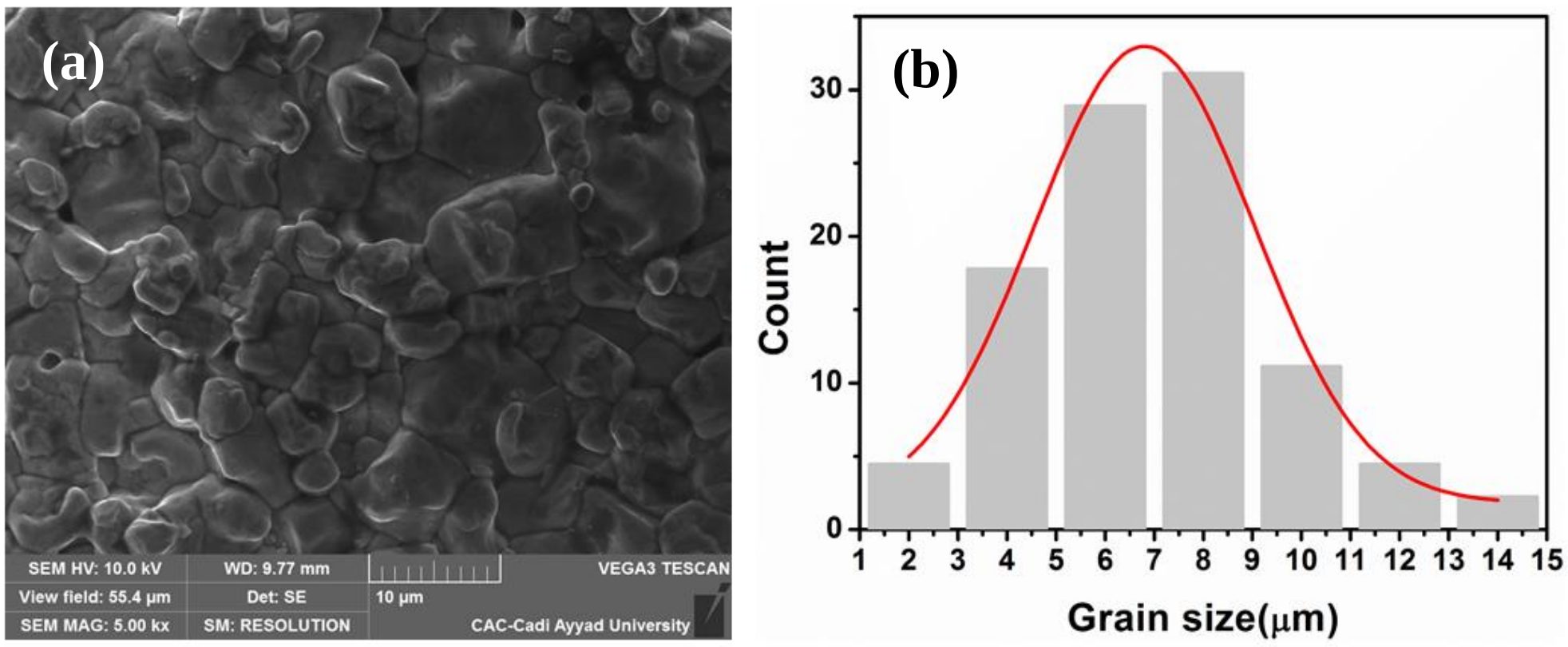
(a)
SEM HV: 10.0 kV
WD: 9.77 mm
VEGA3 TESCAN
View field: 55.4 µm
Det: SE
10 µm
SEM MAG: 5.00 kx
SM: RESOLUTION
CAC-Cadi Ayyad University
(b)
Count
Grain size(µm)

Figure 2

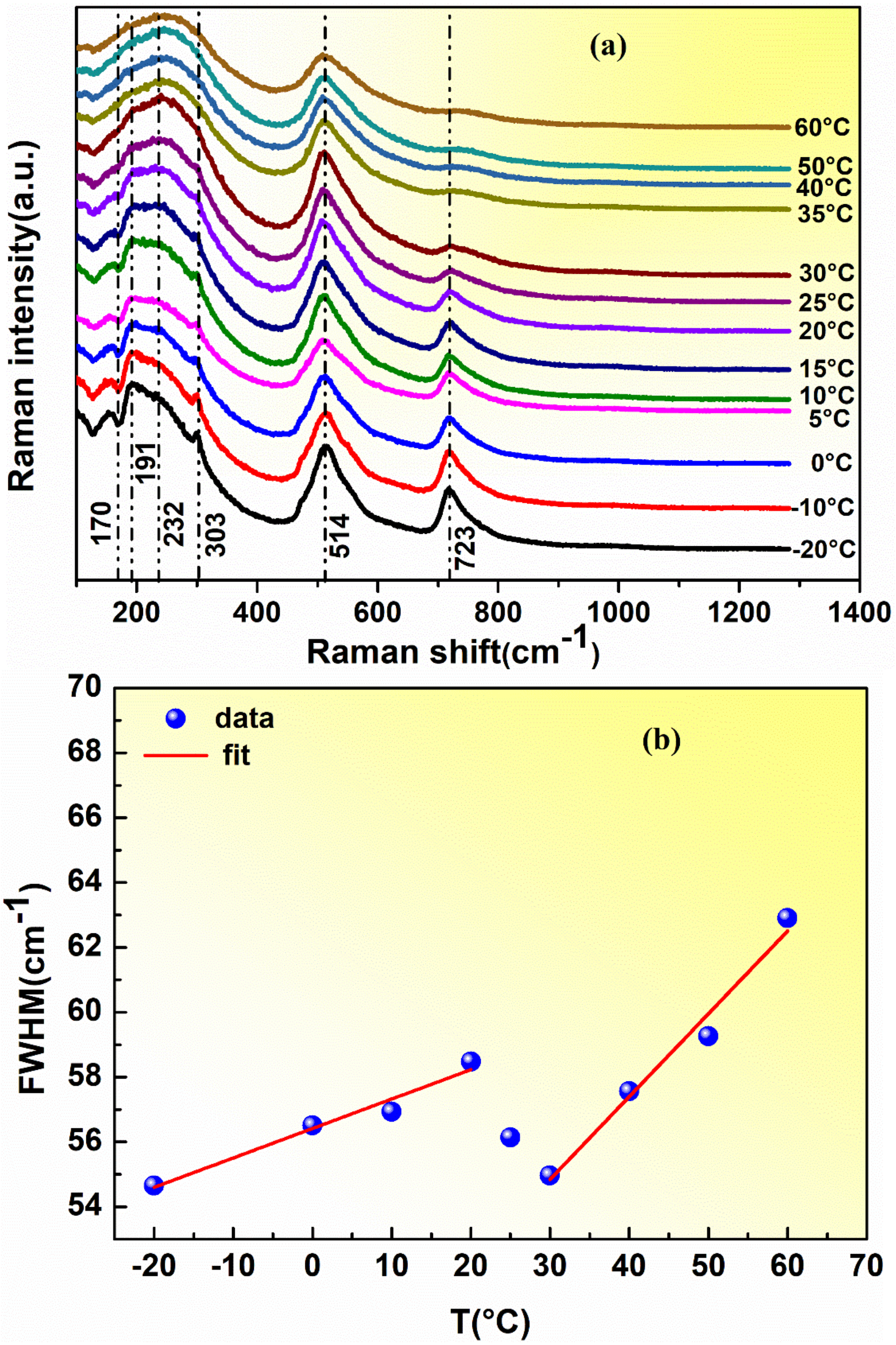

Figure 3

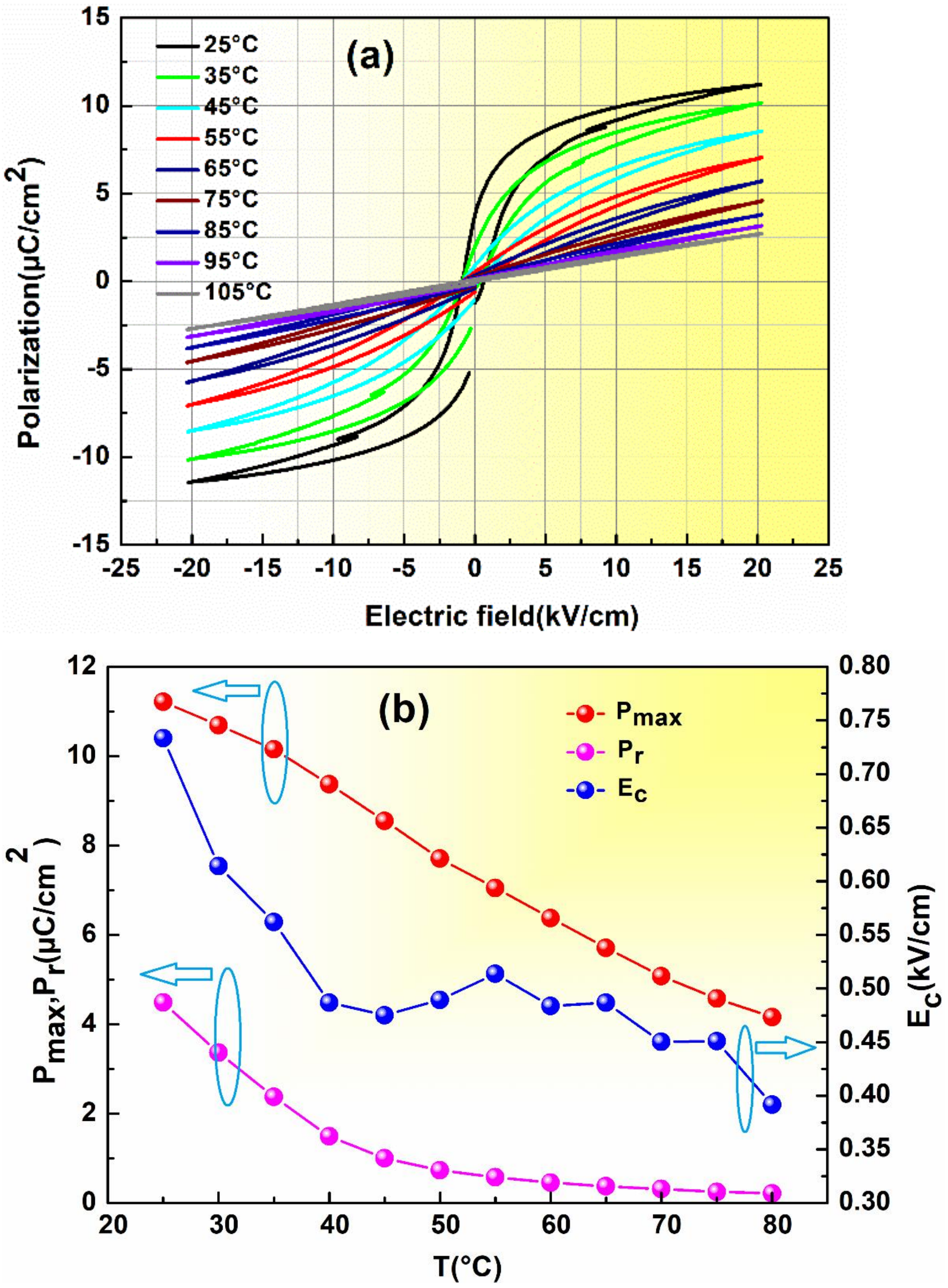

Figure 4

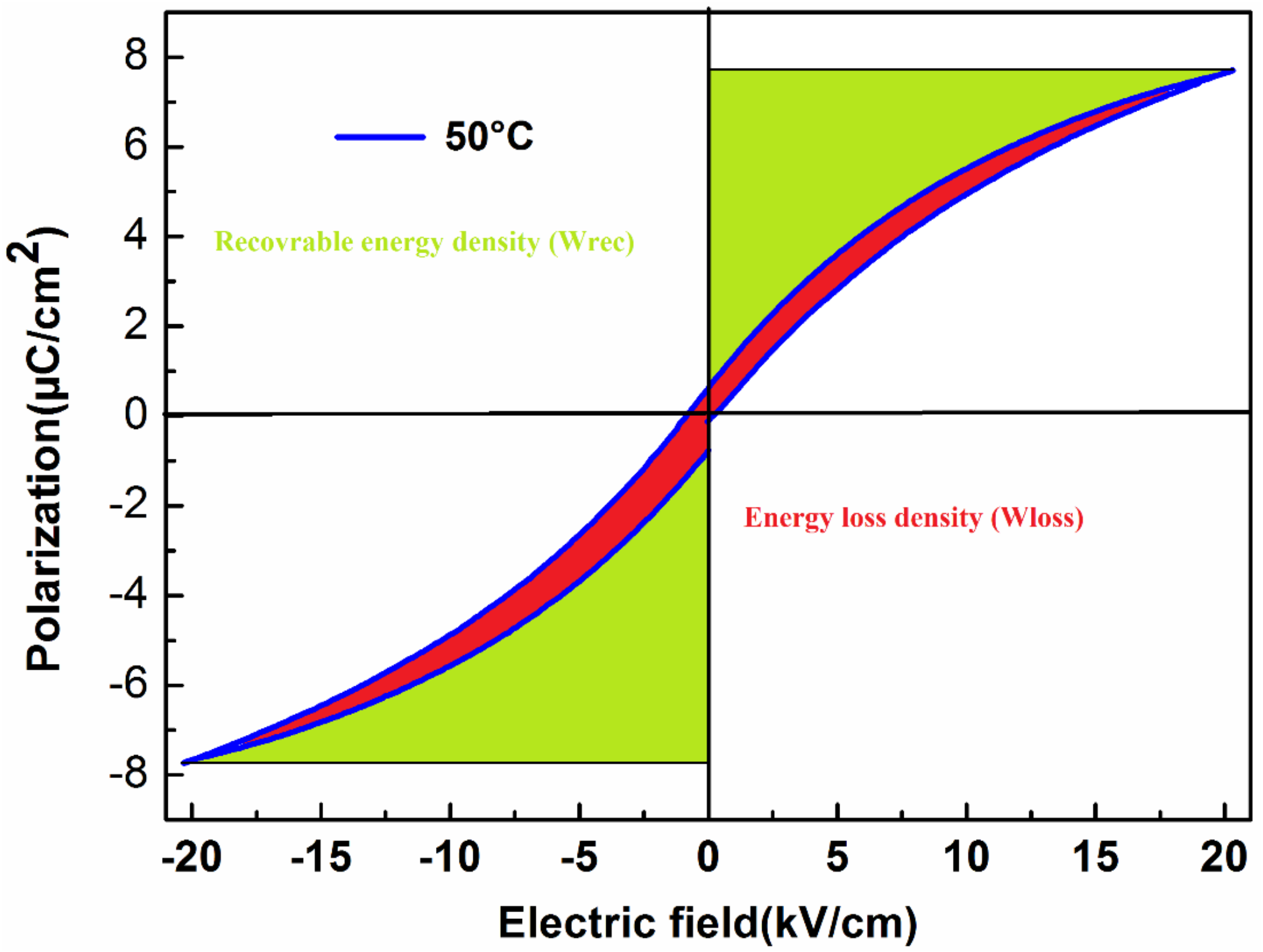

Figure 5

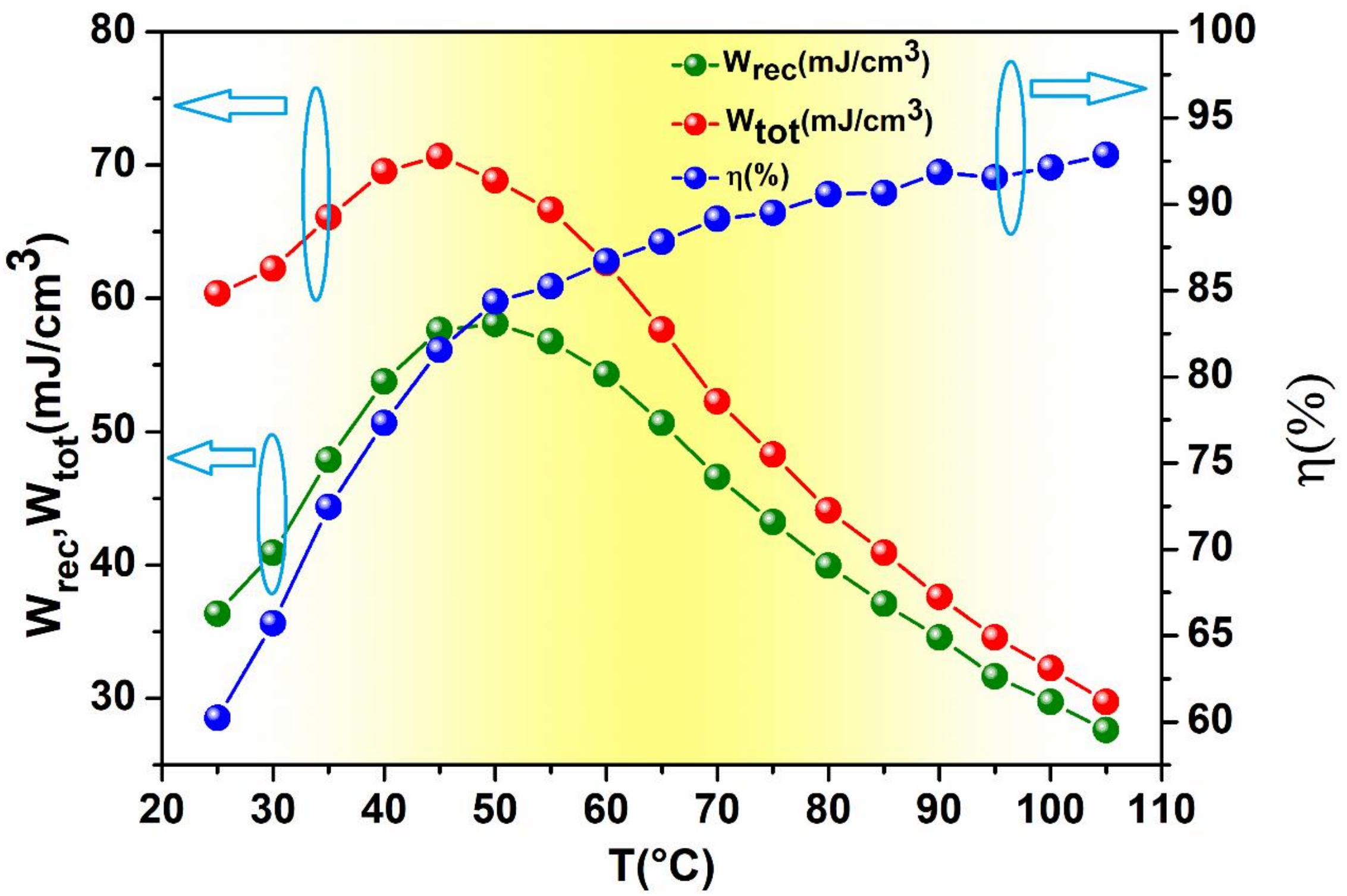

Figure 6

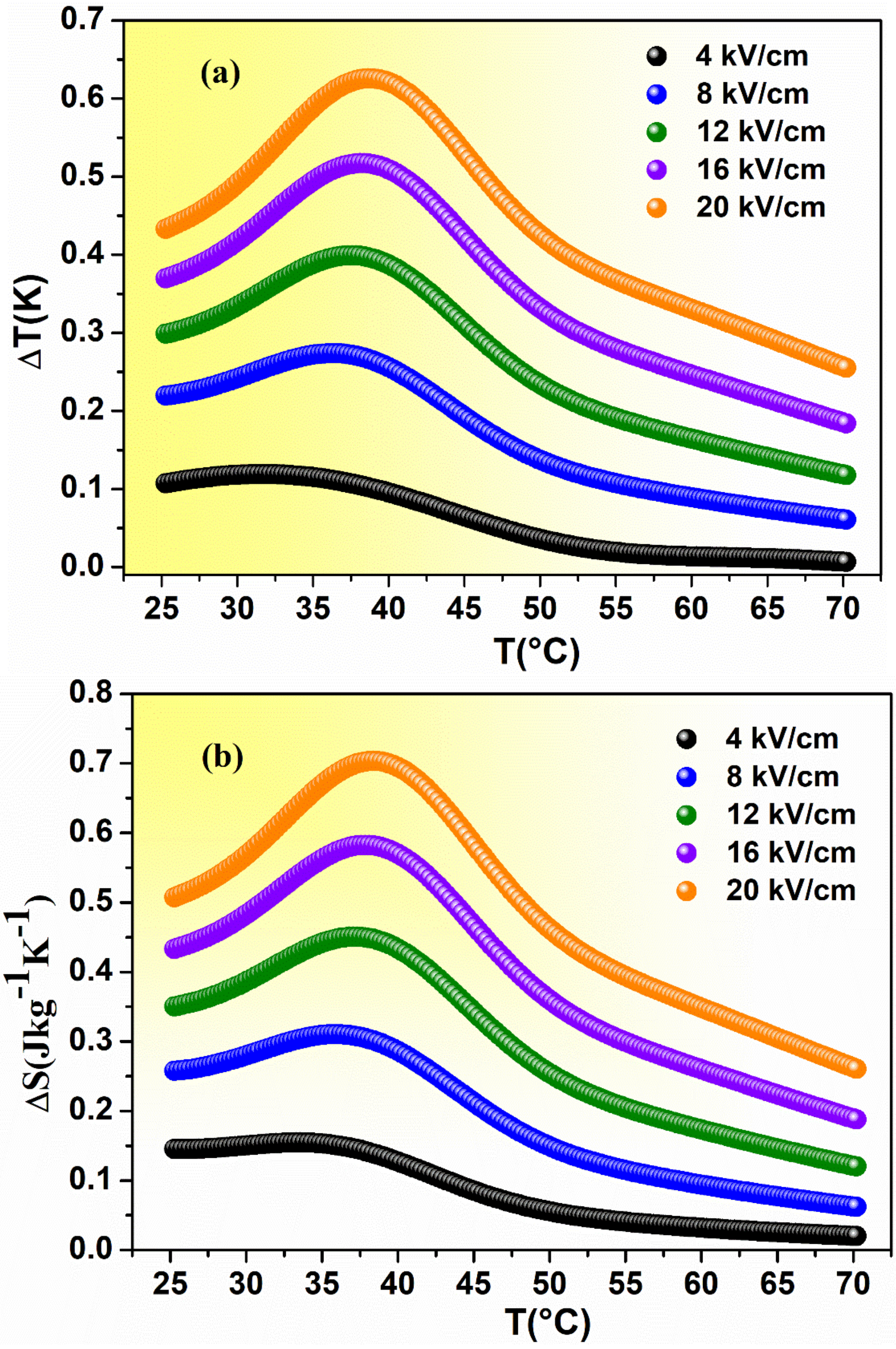

Figure 7

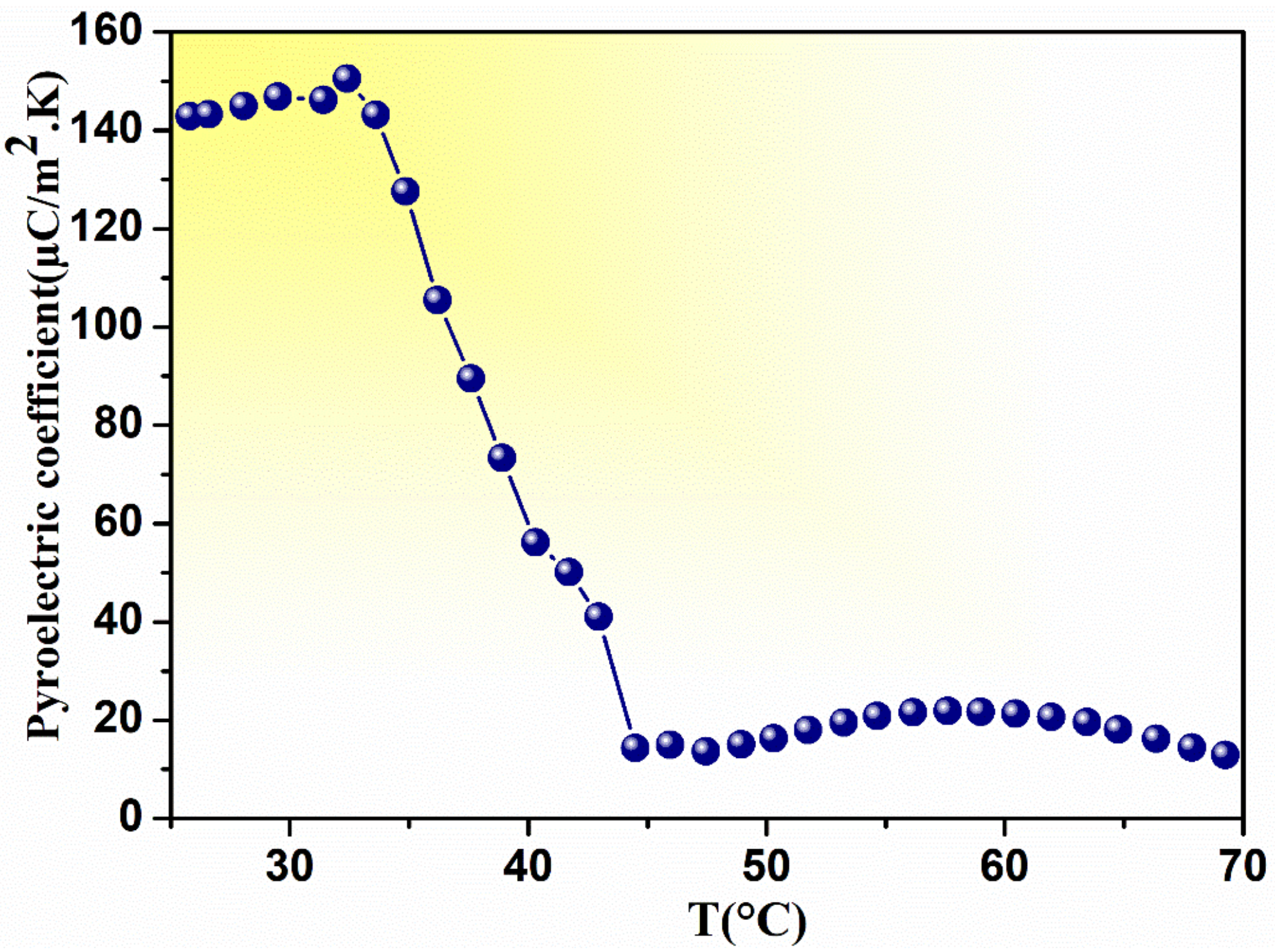

Figure 8

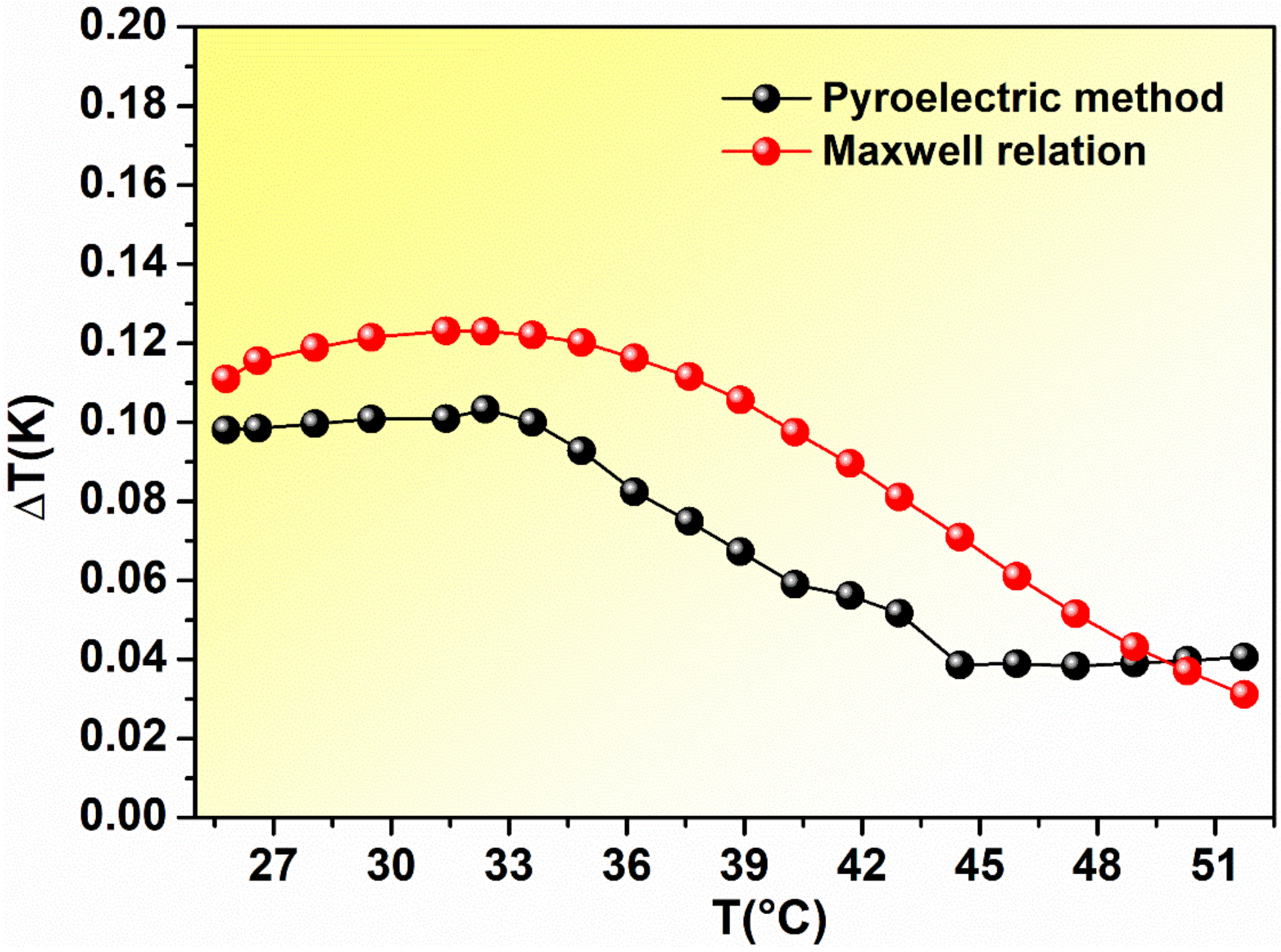

Figure 9